# Thermoelectric properties of high quality nanostructured Ge:Mn thin films

D. Taïnoff[2*], A. Barski[2], E. Prestat[2], D. Bourgault[2], E Hadji[2], Y. Liu[2], P. Bayle-Guillemaud[2], O. Bourgeois[1]*

1-Institut Néel, CNRS and UJF, UPR 2940, 25 rue des martyrs, 38052 Grenoble, France

2- CEA/INAC/SP2M Grenoble, 17 rue des martyrs, 38052 Grenoble, France

We report on the elaboration of germanium manganese nanostructured thin films and the measurement of their thermoelectric properties. We investigate the growth of Ge:Mn layers along with a thorough structural characterization of this materials at the nanoscale. The room temperature thermoelectric properties of these layers containing spherical inclusions are discussed regarding their potential as a model of "electron crystal phonon glass material". We show that the thermal conductivity can be decreased by a factor of 30, even if the electronic properties can be conserved as in the bulk. The thermoelectric performance ZT of such material is as high as 0.15 making them a promising thermoelectric p-type material for Ge related application.

*: Author to whom the correspondence should be sent

dimitri.tainoff@grenoble.cnrs.fr

olivier.bourgeois@grenoble.cnrs.fr

Since the beginning of the 90's and the work of Dresselhaus et al.[1] a lot of works have been focused on improving of the thermoelectric efficiency using nanostructuration[2]. The structuration at the nanoscale may improve the electronic properties by introducing DOS modification or degrade the thermal properties by inducing phonon diffusion. The efficiency of thermoelectric materials is given by the dimensionless figure of merit $ZT = \frac{S^2 \sigma T}{k}$ where S is the Seebeck coefficient, σ the electrical conductivity and k the thermal conductivity. By playing either on the electrical or on the thermal properties, the nanostructuration can strongly increase that factor. In most cases, the increase of the thermoelectric efficiency is obtained thanks to the difference of mean free path existing between phonons (~ 100 nm at 300 K) and electrons (~ 1 nm at 300 K). By introducing, in a given material, a structural disorder at the nanometer scale, it is possible to induce phonon diffusion without affecting the charge carriers justifying the designation of "electron crystal - phonon glass" material.

Calculating the electronic and the thermal properties of an actual "electron crystal - phonon glass" material is a very difficult task. Indeed, the electronic properties of a nanostructured thermoelectric material are never those of the related bulk ones. Different phenomena affecting thermoelectric properties occur concomitantly in most of the real material like the diffusion of both carriers and phonons induced by the nanostructuration[3], the mixing between phononic effect and the diffusion effect in heterostructures[4]. Moreover, the differences in the surface morphology, interface quality, grain boundary[5] or defects will also perturbe the thermoelectric properties.

The elaboration of model material is thus of prime importance to study the physical mechanisms responsible for the increase of figure of merit in the nanostructured material.

---

[1] Hicks and Dresselhaus, *Phys.Rev.B.*, **47,** 16631 (1993)
[2] M.S. Dresselhaus *et al. Advanced Materials* **19**, 1 (2007) ; G.J., Snyder and E.S., Toberer, *Nature materials* **7**, 105 (2008)
[3] Woochul Kim PhD
[4] S. T. Huxtable, A. R. Abramson, C.-L Tien, A Majumdar et al. Appl. Phys. Lett. 80, 1737 (2002)
[5] Poudel et al. *Sciences* ***320***, *634 (2008)*

Such a material should be constituted by a perfectly crystalline part for carrying electrons or holes as well as in the bulk. It must also contain a lot of heterogeneous interfaces for efficient phonon scattering. Thanks to a spinodal decomposition occurring at the nanoscale, germanium manganese thin films is a promising material for reaching these properties[6,7].

The manganese doped germanium thin films were grown by molecular beam epitaxy (MBE) on the (001) face of insulating GaAs substrate. For Mn concentrations ranging from 1 to 10 percents, the low temperature ($T_{gr}$ ~ 100°C) growth of Ge:Mn films give rise to a spinodal decomposition. This decomposition induces the segregation of a manganese rich phase into auto-organized nano-columns embedded in a perfectly crystallized germanium matrix[8,9]. During the growth, a small part of Mn atoms are incorporated within the matrix making the Ge:Mn layers highly p-type[6]. When using well adapted conditions the growth of GeMn thin film is bidimensional and the reflection of high energy electron diffraction (RHEED) pattern of the Ge:Mn layers exhibit a 2x1 reconstruction as shown in figure 1.a).

This RHEED pattern changes drastically after in-situ annealing of the Ge:Mn films at temperature higher than 300°C. At $T_{annealing}$ ~ 300 °C, the spinodal decomposition precipitates causing a strong disorder and the apparition of several other phases near the surface. These lattices cause the apparition of lot a spot are shown on the RHEED pattern of the figure 1.b). When increasing the annealing temperature to $T_{annealing}$ ~ 500 °C the RHEED show only one other lattice over the germanium. This lattice could correspond to a $Ge_3Mn_5$ stable phase. Finally, for temperature increasing up to $T_{annealing}$ ~ 800 °C the RHEED exhibit again 2x1 reconstruction characteristic of a perfect Ge surface.

---

TEM investigations have been carried out in order to reveal the morphology of the layers. They show that the manganese rich nano-columns induced by the spinodal decomposition precipitate during the in-situ post growth annealing to form nearly spherical nano-inclusions. The size and concentration of inclusions depend both on the size/concentration of the nano-column and the subsequent annealing temperature. This can be seen on figure 2.a) and 2.b) for two layers having the same Mn concentration but annealed respectively at 500°C (figure 2.a) and 800°C (figure 2.b). High resolution TEM images (figure 2.c) as well as diffraction cliché confirm that the nano-inclusions are crystallized in the stable $Ge_3Mn_5$ phase as expected from the phase diagram and the RHEED pattern. This phase is metallic and crystallizes preferentially with their c axis along the (001) axis of the Ge[7]. Nevertheless the precipitation of manganese rich $Ge_xMn_y$ nano-columns into well defined $Ge_3Mn_5$ spherical inclusions does not induce extended defect in the germanium matrix. In addition this morphology can be reproduced for at least 1.2 μm thick layers using germanium substrate (see add. material).

In the figure 2.d), the mean free path of phonons and charge carriers are compared to the length characterizing the Ge layers doped with $Ge_3Mn_5$ nano-inclusions for the case of $Ge_{0.94}Mn_{0.06}$ annealed at 800°C. It appears that phonons might be strongly diffused by the nano-inclusions without affecting the electrical transport. Compared to other nano-inclusions in matrix materials[10] the structural properties of Ge:Mn offers several advantages mainly due to (i) the possibility to regulate the size and the concentration of nano-inclusions on a large scale (ii) an higher size dispersion of inclusion and (iii) a well defined interface between inclusions and Ge matrix. Moreover, taking into account the high Seebeck coefficient of germanium, its high carrier mobility and its compatibility with standard technology, Ge:Mn nanostructured thin films could exhibit very competitive ZT value.

---

[10] Kim, W. et al.. *Phys. Rev. Lett.* **96**, 045901 (2006).

We now focus on the study of the thermoelectric properties of Ge:Mn material. We have first measured the electrical conductivity and the Seebeck coefficient of Ge:Mn layers. Measurements were carried out using a home-made experimental setup at room temperature[11] and summarized on the table 1 for different elaboration conditions. The positive sign of Seebeck coefficient first confirms that all Ge:Mn layers are p-type. Seebeck values show no drastic decrease of the Seebeck coefficient due to the presence of the nano-inclusions. Indeed the value of the Seebeck coefficient of Ge:Mn are in the range of 150 to 180 µV which corresponds to standard value for highly p-doped germanium material ($p \sim 10^{19}$ cm$^{-3}$) and confirms previous measurements[12].

Electrical conductivities vary from ~ 1000 $\Omega^{-1}.m^{-1}$ to 20000 $\Omega^{-1}.m^{-1}$. The lowest values of electrical conductivity are obtained both for the non intentionally doped Ge layer and the Ge:Mn layers annealed at 500°C. The similarity of these values is tricky. Indeed the high value of electrical conductivity measured in the n.i.d. Ge layer is due to a concentration of residual p-type dopant incorporated during the growth. For the Ge:Mn layers annealed at 500°C, the low value of the electrical conductivity is due to a higher concentration of carriers and thus a lower mobility, as previously observed[12]. This low mobility can be explained considering the figure 2.a). Indeed the mean spacing between the inclusions is similar to the mean free path of carriers. This induces carrier diffusion and thus decreases the electrical conductivity. In order to reach the properties of an "electron crystal – phonon glass" material and thus take advantage of the high mobility of carrier in Ge, it is necessary to lower the concentration of inclusions. The value of electrical conductivity was thus measured for the same Mn concentration in a layer annealed at 800°C corresponding to a lower concentration of inclusions (cf figure 2.b). Results show an increase of more than one order of magnitude as compared to other Ge:Mn layers. This is explained by the mean spacing between the nano-inclusions which is now more than 100 times bigger than the carrier mean free path. In that case the electrical conductivity corresponds to a highly p type Ge bulk material doped with *p* ~ 3-5 $10^{19}$ cm$^{-3}$ [13,14] which is coherent with the Seebeck value measured in bulk material. That shows that in this case the couple size/concentration of inclusions is well choiced to strictly reach the properties of an "electron crystal" bulk material.

---

[11] D. Bourgault, C. Giroud Garampon' N. Caillault' L. Carbone' J.A. Aymami' *Thin Solid Films* **516**, 8579 (2008)
[12] Yu Ing Song PhD thesis

In order to estimate their potential as a model of "electron crystal –phonon glass" material, the thermal conductivity of the Ge:Mn layers was computed using the modified Callaway approach developed by Kim et Majumdar[15,16]. This modeling was carried out including the contribution of both transverse and longitudinal modes for different size/concentration/dispersion of inclusions[17]. Size distribution of nano-inclusions was deduced from the TEM measurements and their concentration was estimated from both plane and cross TEM views by making the approximation of spherical inclusions. Results are displayed on figure 3 for bulk Ge (black curve) and Ge$_{0.94}$Mn$_{0.06}$ layer annealed at 800° (red curve). In that case the thermal conductivity at room temperature is reduced by factor 30 as compared to bulk Ge showing that Ge:Mn thin films are also very good "phonon glass" material. The value of ZT calculated from the measurements of electrical conductivity and Seebeck coefficient and the thermal conductivity modeling is as high as 0.15 which is at the state of the art for column IV semiconductor material and could be improved by introducing an alloy disorder using Ge isoelectronic element i.e. Si and/or Sn. Finally Ge:Mn layers appear as a model electron crystal phonon glass material since their electrical properties are fully related to the Ge matrix while the thermal transport is strongly limited by inclusions.

This experimental work on a model material illustrates the importance of the electrical transport properties which might not be sacrificed for reducing thermal conductivity.

To conclude we have demonstrated the growth of germanium manganese thin layers which can serve as a model of "electron crystal phonon glass" material. The existence of a spinodal decomposition in this system allows the formation of well defined $Ge_3Mn_5$ nano-inclusions in a perfect highly p-doped Ge matrix. The measurement of Seebeck coefficient and electrical conductivity of samples annealed at different temperature has shown that the electrical properties of Ge:Mn layers can reach those of highly doped bulk materials. On the other hand the modeling of Ge:Mn thermal transport properties shows that thanks to a good size/concentration/dispersion the room temperature Ge:Mn thermal conductivity can be reduced by a factor 30 while keeping the bulk electrical conductivity. A ZT value as high as 0.15 is deduced for Ge:Mn layer making it a promising p-type thermoelectric material.

This work is partially supported by the European FP7 program MERGING project (grant agreement no 309150)

**Figure Caption**

**Figure 1**: RHEED images recorded during the elaboration of the Ge3Mn5 nanoinclusions : a) After the low temperature growth of the Ge:Mn layers, the RHEED is sticky and exhibit the 2x1 reconstruction; b) After annealing at 300°C the RHEED show several features due to the disorder induces by the precipation of the nanocolumns. c) When increasing the temperature at 500°C the RHEED is characteristic of a Ge surface. The spot near the first diffraction stripe of Ge lattice corresponds to the Ge3Mn5 lattice parameter; d) At 800°C the RHEED exhibit again the 2x1 reconstruction. No traces of inclusions are visible in the RHEED images.

**Figure 2**: TEM images of $Ge_{0.94}Mn_{0.06}$ layers annealed at a) 500°C and b) 800°C; c) high resolution image of $Ge_{0.94}Mn_{0.06}$ layers annealed at 800°C showing the high crystalline quality of the Ge matrix despite of the lattice parameter difference between the $Ge_3Mn_5$ inclusions and the Ge matrix; d) mean free paths are indicated for phonons and carriers for a $Ge_{0.94}Mn_{0.06}$ layers annealed at 800°C.

**Figure 3**: Thermal conductivity of Ge:Mn layers containing Ge3Mn5 nano-inclusions modeled for size/concentration/dispersion extract from TEM study for $Ge_{0.94}Mn_{0.06}$ layers annealed at 800°C.

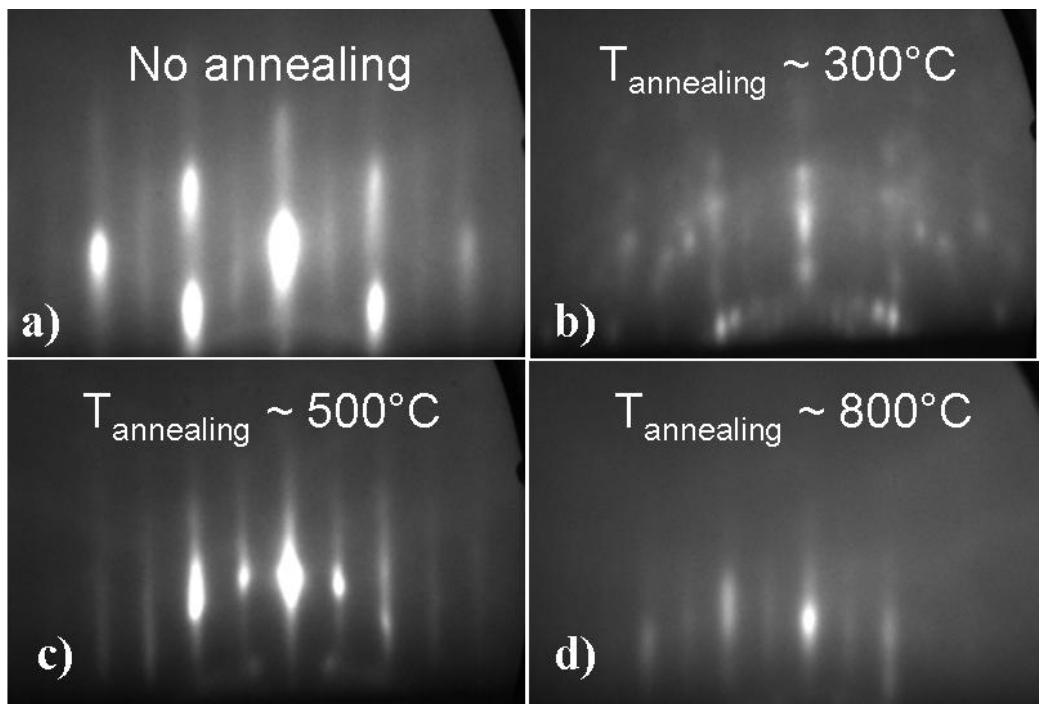

Fig. 1 Tainoff et *al.*

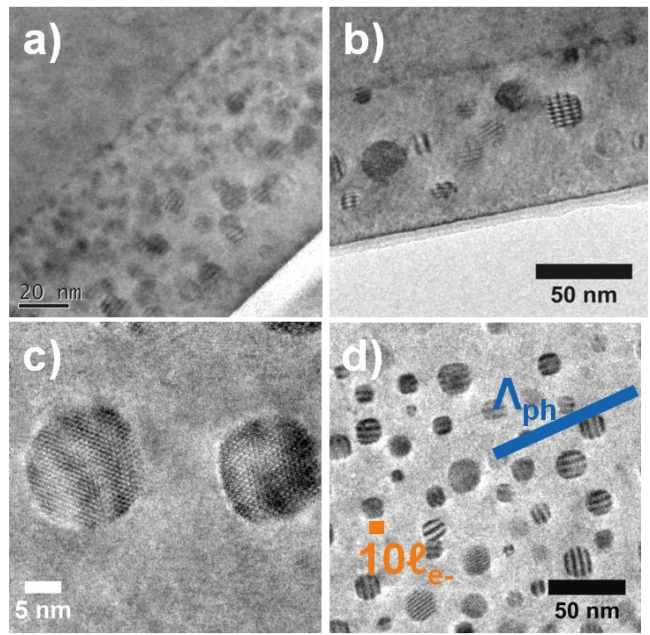

Fig. 2 Tainoff et *al.*

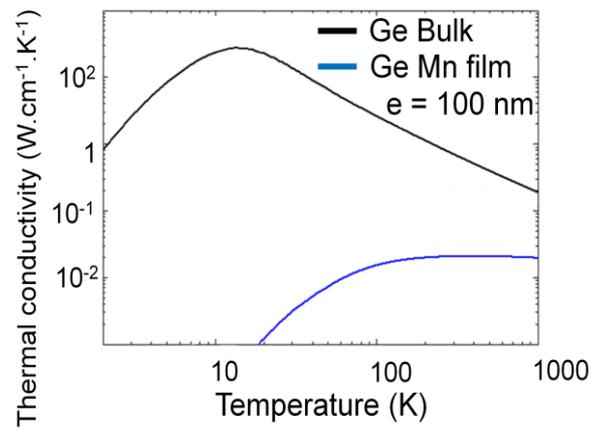

Fig. 3 Tainoff et *al.*

**Table Caption**

**Table 1**: Values of electrical conductivity and Seebeck coefficient measured for Ge:Mn nanostructured thin films elaborated using different growth conditions.

|  | Ge n.i.d. | $Ge_{0.98}Mn_{0.02}$ | $Ge_{0.94}Mn_{0.06}$ | $Ge_{0.94}Mn_{0.06}$ |
|---|---|---|---|---|
| Annealing temperature | no | 500°C | 500 °C | 800°C |
| Growth temperature | 150°C | 150°C | 150°C | 150°C |
| Conductivity ($\Omega^{-1}m^{-1}$) | 1000 | 800 | 1100 | 20000 |
| Seebeck ($\mu V.K^{-1}$) | 200 +/- 10 | 180 +/- 10 | 150 +/- 10 | 150 +/- 10 |

Table 1 Tainoff et *al.*